\documentclass{tCPH2e}
\begin{document}
\title{Ion Coulomb Crystals}
\author{ Richard C. Thompson\\
Imperial College London, London SW7 2AZ, UK}

\date{\today}

\maketitle
\begin{abstract}
Ion Coulomb crystals (ICC), formed by atomic ions at low temperatures in radiofrequency and Penning ion traps, are structures that have remarkable properties and many applications.  Images of Coulomb crystals are striking and reveal the crystal structure, which arises from a balance between the trapping forces acting on the ions and their mutual Coulomb repulsion.  Applications of these structures range from frequency standards and quantum simulation through to measurement of the cross sections of chemical reactions of ions.

\end{abstract}%

\section{Introduction}

Many solids exist in the form of a crystal, where atoms are arranged in a regular pattern and have fixed positions relative to each other.  Such crystals are normally three-dimensional but atoms can also, in some cases, form two-dimensional structures such as graphene.  The equilibrium distances between atoms are essentially determined by the overlap of the atomic wavefunctions, resulting in typical inter-atomic distances of 0.3\,nm and a density of the order of 3$\times10^{28}$  m$^{-3}$.  When it is heated, a crystal will often undergo a transition to a liquid, in which the density remains similar to that of the crystal but the atoms no longer have fixed positions relative to each other.

Atomic ions can also form crystal structures when they are held in ion traps.  Ion traps are devices that use a combination of electric and magnetic fields to confine the motion of charged particles (in this case, atomic ions) to a small region of space.   When the ions have low energies, they accumulate at the centre of the trap, but since they repel each other due to the Coulomb force, there is a limit to how close they can approach each other. In equilibrium, and at low enough temperatures, the ions will then form a regular crystal-like structure with the inter-particle spacing determined by the balance between the trapping fields and the  Coulomb repulsion between the ions. For typical values of the trapping fields (as discussed below) this results in a spacing of the order of 10\,$\mu$m and hence a density of the order of $10^{15}$  m$^{-3}$.  This is many orders of magnitude lower than the density of conventional crystals and even of the air around us.  In fact, it is comparable to the density of residual gas molecules in an ultra-high vacuum system (such as would be used to enclose the ion trap).  These novel crystal structures therefore have many interesting properties that contrast with those of conventional crystals.  On the other hand, there are also some similarities of behaviour, even though the densities differ by many orders of magnitude.

This review discusses the formation, structure and applications of ion Coulomb crystals (ICC).  It is arranged as follows.  Section \ref{sec:traps} presents the basic physics of ion traps and cooling techniques; Section \ref{sec:1DICC} discusses the properties of one-dimensional crystals (i.e. strings or chains of ions) while Section \ref{sec:3DICC} deals with the three-dimensional case.  Section \ref{sec:applications} discusses applications including experiments to study phase transitions and quantum mechanical effects in ion Coulomb crystals.  Finally, Section \ref{sec:conclusions} presents some conclusions.

\section{Ion traps and laser cooling} \label{sec:traps}

\subsection{The linear RF trap}  

Ion traps were first developed in the late 1950s and the three-dimensional radiofrequency  trap, often called the Paul trap, was developed in Bonn by the group of Wolfgang Paul, who was  awarded a Nobel prize in 1989 for his work in this area \cite{rftrap,paulnobel}.  There are many reviews of ion traps available which explain in detail how the radiofrequency trap works and discuss the many applications that have been developed:  see for example \cite{thompson1993,CPHorvath,CPozeri,CPcoldatoms}.

\begin{figure}
\begin{centering}
 \includegraphics[width=50mm]{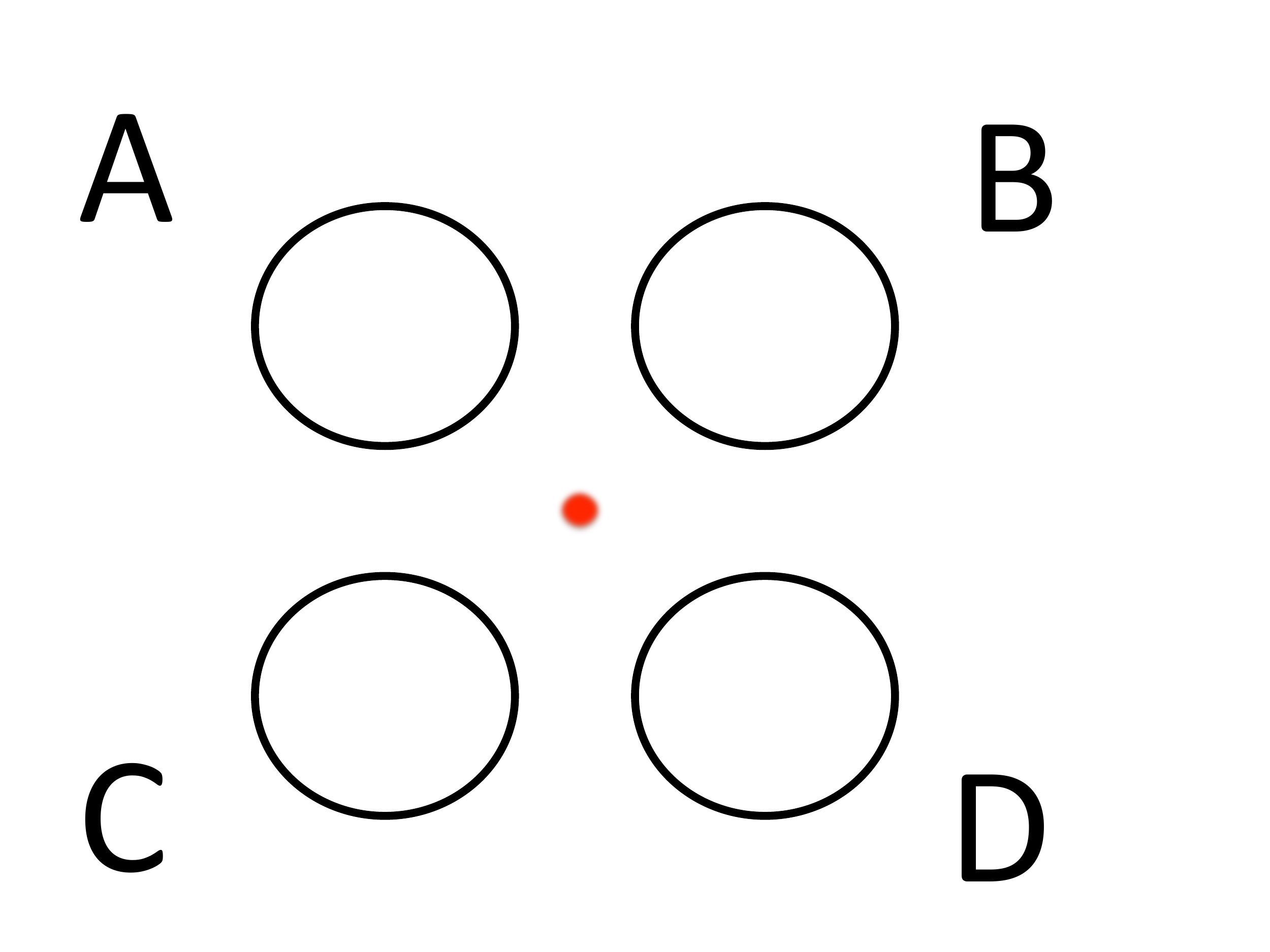}
\caption{Cross-section through the electrodes of a linear radiofrequency trap constructed from cylindrical rods.  Electrodes A and D are connected together, as are electrodes B and C.  A static (DC) voltage applied between the two pairs of electrodes creates a two-dimensional quadrupole  potential in the region between the electrodes, with a saddle point at the centre.  As discussed in the text, an oscillating voltage can  give rise to a two-dimensional pseudopotential which traps charged particles along the axis of the trap (into the page). The distance from the centre of the trap to each electrode ($r_0$) is typically of the order of 1\ mm (see text).}
\label{fig:linear}
\end{centering}
\end{figure}

In this review, we will concentrate mainly on a two-dimensional version of the Paul trap, called the linear radiofrequency (RF) trap. Briefly, the linear RF trap  works by using a set of four electrodes (usually made from circular rods: see Figure \ref{fig:linear}) to generate a quadrupole potential in a region of space under ultra-high vacuum conditions.  The rods are aligned parallel to the $z$-axis and the radiofrequency potential is required in order to confine ions in the $xy$-plane, i.e. to push them towards the central $z$-axis.  

It is not possible to achieve this radial confinement by applying a static (DC) potential.  To see this, consider the effect of applying a positive potential to two rods opposite each other, with the other two electrodes grounded.  This would tend to push a positively-charged ion away from the positively charged rods and towards the $z$-axis, but the ion would also be attracted to the two other rods.  This would lead to the loss of the ion.  So with a static potential it would be possible to trap in (say) the $x$-direction but only with the motion in the orthogonal $y$-direction being unstable.  Reversing the sign of the applied voltage would have the opposite effect, giving stable motion along $y$ and unstable motion along $x$.  However, it turns out that if the applied potential is made to oscillate between positive and negative values, such that the stable and unstable directions are constantly reversing, the final time-averaged motion of the ion can be stable in \emph{both} directions  simultaneously, so long as the oscillation frequency is set to be within a certain range \cite{douglasreview}.  

The average applied potential in this situation is zero, so it is not intuitively obvious that stable trapping can be achieved, but this can be demonstrated from a careful mathematical analysis of the system \cite{douglasreview}.  This analysis shows that the final motion consists of an oscillation of the ion in the plane perpendicular to the $z$-axis of the trap as if it were sitting in an effective potential that had a two-dimensional minimum along this axis. This is called the  \emph{pseudopotential} and this motion is referred to as the \emph{secular motion}.   Superimposed on this motion is a faster and smaller-amplitude oscillation at the  frequency of the applied field, called the \emph{micromotion}.  The pseudopotential arises because this driven micromotion takes place in a nonuniform field such that although the  electric field at any point averages to zero,  the average force acting on the ion is non-zero because it is moving around inside  the trap. 

The motion is properly described by a Mathieu equation and solution of this equation gives an expression for the oscillation frequencies\footnote{Note that throughout this article we will refer to oscillation frequencies using the angular frequency $\omega$  but where we give a numerical value for the frequency we will quote the value of $\omega /2\pi$ in Hz.}: 
\begin{equation}
\omega_x^2=\omega_y^2= (q^2/2)(\Omega/2)^2
\label{eq:omega}
\end{equation}
where $\Omega$ is the angular frequency of the applied RF potential (with amplitude $V$) and $q$ is given by
\begin{equation}
q\approx 4eV/m\Omega^2r_0^2.
\label{eq:q}
\end{equation}
Here $r_0$ is the internal radius of the trap; $e$ and $m$ are the charge and mass of the ion respectively. Equation \ref{eq:omega} is valid for $q\ll 1$. Equation \ref{eq:q} is exact if the electrodes have a hyperbolic rather than a circular cross-section.  Circular electrodes do not produce a pure quadrupole potential like hyperbolic ones do, but they are easier to manufacture.  The additional higher order terms they produce in the potential are not important close to the trap axis.   

It is also necessary to confine the particle's motion in the $z$-direction, i.e. along the axis of the trap, and this is achieved  by applying an additional static potential at the two ends of the trap.  This is sometimes done by segmenting the RF rods and adding a DC potential to the end segments, and sometimes it is done by including extra \emph{endcap} electrodes at the ends of the rods.  In either case the result is a harmonic potential along the axis that pushes ions back towards the central part of the trap.  This potential is usually relatively weak, with the confinement in the radial direction (from the RF pseudopotential) much stronger.  This in turn means that the ion oscillation frequencies are generally higher for the transverse ($xy$) motion and lower for the axial ($z$) motion.

Linear RF traps can vary in size but for the experiments to be described here, typical dimensions involve a separation of the rod electrodes on the order of 1\,mm with the trapping region a few mm in length.  The applied frequency would be typically 1 to 10\,MHz and the resulting secular oscillation frequencies would be in the region of a few hundred kHz.  However, some traps may be several mm across and others, manufactured using microfabrication techniques, are as small as 100\,$\mu$m and typically have much higher oscillation frequencies \cite{CPHensinger}.

We have described the trapping mechanism for a single ion but the same principles also apply when many particles are being trapped.  However, it is clear that although all ions will feel a force towards the centre of the trap, eventually there must come a point at which the Coulomb repulsion between the ions balances the trapping potential.  This leads to a maximum density of particles that can be accommodated in the trap.  The effect can be described in terms of the space charge potential that builds up as the particles accumulate.  It can be shown that space charge effects lead to a uniform density of particles in the trap at low temperatures, which has important consequences for large ion Coulomb crystals.  

\subsection{Radiofrequency (RF) micromotion}

An important feature of all radiofrequency ion traps is the presence of micromotion.  This is the small-amplitude motion of ions which is driven directly by the applied radiofrequency field.  Since the pseudopotential arises as a result of this motion, it is an unavoidable feature of RF traps. It is important because it can lead to heating up of an ion cloud (in effect, it can couple the energy stored in the RF field into the secular motion of the ions).  It will also lead to Doppler broadening effects in the spectra of ions in traps.  For these reasons, it is often desirable to minimise or eliminate the micromotion. In the linear trap, the RF field is confined to a plane perpendicular to the $z$-axis. Since the field approximates to a quadrupole field (remember it is generated by four electrodes), the amplitude of the field (and the resulting micromotion) rises linearly with the distance from the axis of the trap.  This means that along the axis of the trap (i.e. along the line $x=y=0$) there is no micromotion.  This is a point to which we will return later.

\subsection{The Penning trap}

\begin{figure}
\begin{centering}
 \includegraphics[width=80mm]{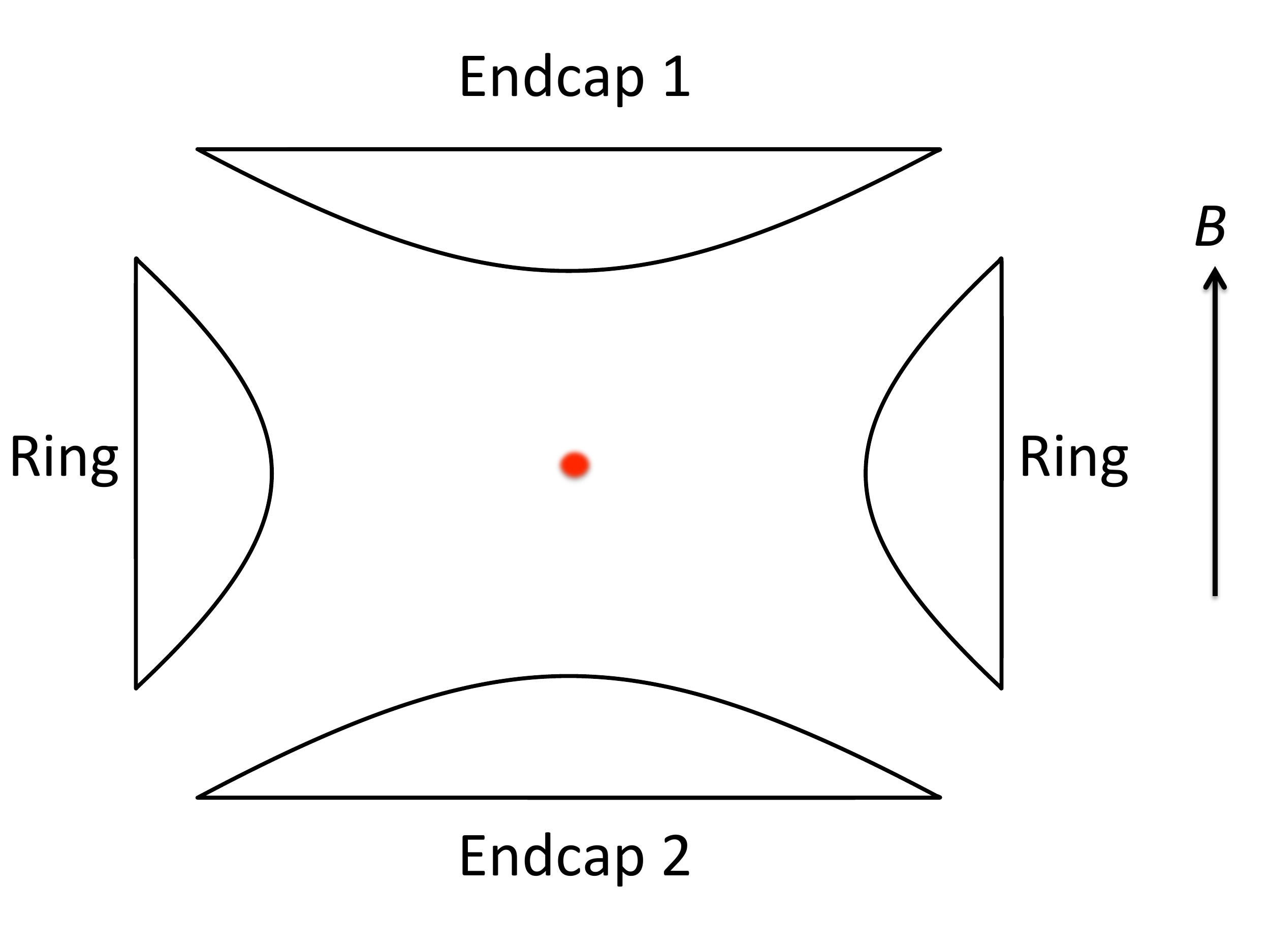}
\caption{Cross-section through the electrodes of a Penning trap.  The two endcap electrodes are connected together.   A DC potential applied between the endcaps and the ring creates a three-dimensional quadrupole potential in the region between the electrodes. This traps particles along the vertical $z$-direction.  The addition of a uniform magnetic field $B$ along the trap axis ensures three-dimensional trapping.  The separation of the endcaps is $2z_0$ and the internal diameter of the ring is $2r_0$, which is typically of the order of 1\ cm (see text).}
\label{fig:trap}
\end{centering}
\end{figure}

The Penning trap offers an alternative way of confining the motion of ions.  In this type of trap, three-dimensional confinement is achieved by using three electrodes (a ring and two endcaps) having the shape of hyperboloids of revolution about the $z$-axis (see Figure \ref{fig:trap}).  The hyperbolic shape of the electrodes results in a quadrupole electrostatic saddle potential given by
\begin{equation}
\phi(r,z) = U_0 {{2z^2-r^2}\over{2z_0^2+r_0^2}}
\label{eq:saddle}
\end{equation}
where $U_0$ is the applied potential, $2z_0$ is the separation of the endcaps and $2r_0$ is the internal diameter of the ring electrode.   A positive potential $U_0$ between the endcaps and the ring confines the axial motion of a positively-charged ion because the ion is repelled from both endcaps.  The quadrupole potential well therefore gives rise to an axial oscillation of the ion at a frequency $\omega_z$.  The radial motion is unstable because the ion will be attracted towards the ring electrode, which encircles the $z$-axis.  However, the addition of a strong axial magnetic field $B$ stops the ion moving towards the ring and instead forces the motion into a combination of circular orbits in the radial plane (a slow magnetron motion at $\omega_m$ combined with a faster modified cyclotron motion at $\omega_c^{\prime}$).  In this way, three-dimensional confinement is achieved \cite{brown}.  The three trap oscillation frequencies are given by
\begin{equation}
\omega_z^2={{4eU_0}\over{2z_0^2+r_0^2}}
\end{equation}
\begin{equation}
\omega_c^{\prime}=\omega_c/2+\sqrt{\omega_c^2/4-\omega_z^2/2}
\end{equation}
\begin{equation}
\omega_m=\omega_c/2-\sqrt{\omega_c^2/4-\omega_z^2/2}
\end{equation}
where $\omega_c$, the pure cyclotron frequency, is equal to $eB/m$.

One advantage of the Penning trap is that there is no micromotion; however, the energy associated with the magnetron motion is negative, due to the negative radial potential energy described by Equation \ref{eq:saddle}, and this results in complications for the stability of ions in the trap and the effectiveness of laser cooling, as will be discussed below.

The internal diameter of a Penning trap ranges from a few mm to a few cm, depending on the application.  As a very rough guide, a magnetic field of a few tesla (usually from a superconducting magnet) gives typical oscillation frequencies in the range 100\,kHz to 1\,MHz for atomic ions of interest. Penning traps are often constructed from a stack of open cylindrical electrodes rather than hyperbolic electrodes as shown in Figure \ref{fig:trap}. This is because cylindrical electrodes are much easier to manufacture, and in addition, good optical access is generally easier to achieve.  It is possible to design such electrodes to eliminate higher-order terms in the potential they produce close to the centre of the trap \cite{brown}.  

\subsection{Laser cooling}

In order to observe crystals in ion traps, it is necessary to cool the ions, i.e. to reduce their kinetic energy \cite{segalPTCP}.  Since the RF micromotion amplitude in a linear trap depends on the amplitude of the secular motion, it is sufficient to remove energy from the secular motion.  Although there are several other methods available for cooling of particles in traps, including resistive cooling \cite{winterselectronicdetection} and buffer gas cooling \cite{buffergascooling}, the method that is most important for ion Coulomb crystals is laser cooling \cite{lasercooling}.  This is because it is the only cooling method that is able to reduce the temperature of ions to low enough values for crystallisation to be observed.

Laser cooling is based on the exchange of momentum between photons from a laser and the ions.  Each time an ion in the trap absorbs a photon from a laser beam tuned close in frequency to a strongly allowed transition  from the ground state to an excited state of  the ion, the ion absorbs the momentum carried by the photon (equal to $h/\lambda$).  In order to make sure that the effect of this is to slow the ion down, it is necessary for absorption to take place only when the ion is moving towards the laser.  This can be done by making use of the Doppler effect, which is the shift in the frequency of the light as seen by the ion, as a result of its motion.  By setting the laser frequency slightly below resonance, the ion will only see the light as resonant if it is moving towards the laser; if it is moving away from the laser, the Doppler shift takes it further out of resonance and no light is absorbed.  

This process, when repeated many thousands of times, is able to slow ions down from the high energies (typically a few electron volt) which they have when they are created inside the trap or when they enter the trap from elsewhere in the apparatus.  However, there is a limit to how low a temperature can be reached in this way.  This limit comes from the fact that the light must be re-radiated from the ion before it can absorb another photon (this takes place on average after one lifetime of the excited atomic state:  for resonance transitions of ions this lifetime is usually of the order of a few ns).  The emission of a photon is in a  random direction, so although the recoil of the ion averages to zero, it gives rise to a random walk in momentum space which results in an average kinetic energy that is determined by the balance between cooling from the laser and heating from the recoil.  This generally corresponds to an equilibrium temperature of around 1~mK for typical ions of interest \cite{thompson1993}.

This technique is called \emph{Doppler cooling} as it depends on the Doppler effect; there are other types of laser cooling that are able to reach lower temperatures.  These are not generally used for work with large ion crystals, but techniques such as Raman cooling \cite{raman} and optical sideband cooling \cite{sideband}  can be applied to small ion strings in a linear RF trap.  This sub-Doppler cooling is essential for much of the research in quantum information processing carried out using ion traps.

Laser cooling can be applied only to a limited number of different ion species.  This is because an ion must have a suitable resonance transition available at a wavelength that can be reached by continuous-wave lasers, and the energy level structure of the ion must be such that the ion can cycle rapidly between its ground state and the excited state reached by absorption of laser light.  This is only possible directly with Mg$^+$, Be$^+$ and Hg$^+$ ions.  However, if an ion has a metastable (long-lived) state to which it can decay from its excited state, it is sometimes possible to use a second laser to bring the ion back up to the excited state, so that it can resume the cooling process.  This allows a number of other  ions to be used for laser cooling experiments, including Ca$^+$, Sr$^+$, Yb$^+$ and Ba$^+$.  The total number of singly-charged ion species that can be laser cooled is about ten \cite{thompson1993}. There are no suitable doubly-charged ions because the laser wavelengths required are too far into the ultraviolet where no tuneable continuous-wave lasers are available.

From the point of view of the physics of ion Coulomb crystals, the species of ion being used to create the crystals is unimportant (so long as it can be laser cooled effectively).  This is because the physics of the crystal depends only on the temperature of the ions (see the next section) and on the Coulomb repulsion between the ions.  At typical distances of several $\mu$m, the detailed electronic structure of the ions is completely irrelevant as they behave like point charges.

\section{One-dimensional ion Coulomb crystals -- strings of ions}  \label{sec:1DICC}

The simplest type of ion Coulomb crystal is a linear string of ions along the axis of an ion trap.  This arises most naturally in the case of a linear RF trap, where the confinement along the axis of the trap is typically much weaker than the radial confinement.  If several cold ions are present in a trap, they therefore tend to line up along the trap axis -- it is this configuration that has the lowest potential energy.  For $N$ ions along the trap axis, the total potential energy of the system is given by
\begin{equation}
E=\sum_i\frac{1}2 m\omega_z^2 z_i^2+\sum_{i>j}\frac{e^2}{4\pi\epsilon_0\vert z_i-z_j\vert}.
\label{eq:energy}
\end{equation} 
The equilibrium positions of the ions can be found by minimising this energy as a function of the $z_i$.  The resulting positions have been tabulated by James \cite{james} and have been confirmed in many experiments.  The length scale of the linear crystals formed in this way can be found by evaluating the equilibrium separation of two ions, $\Delta z$.  This is the separation at which the inward force due to the trapping potential is exactly balanced by the outward repulsion of the ions, i.e.
\begin{equation}
m\omega_z^2 \frac {\Delta z} 2=\frac{e^2}{4\pi\epsilon_0\Delta z^2}
\end{equation} 
giving $\Delta z=10\,\mu$m for calcium ions having an axial oscillation frequency $\omega_z$ of $2\pi\times$500\,kHz.  As the number of ions increases, the separation between adjacent ions at the centre of the string reduces approximately as $N^{-0.559}$ \cite{james}.

\begin{figure}
\begin{centering}
 \includegraphics[width=80mm]{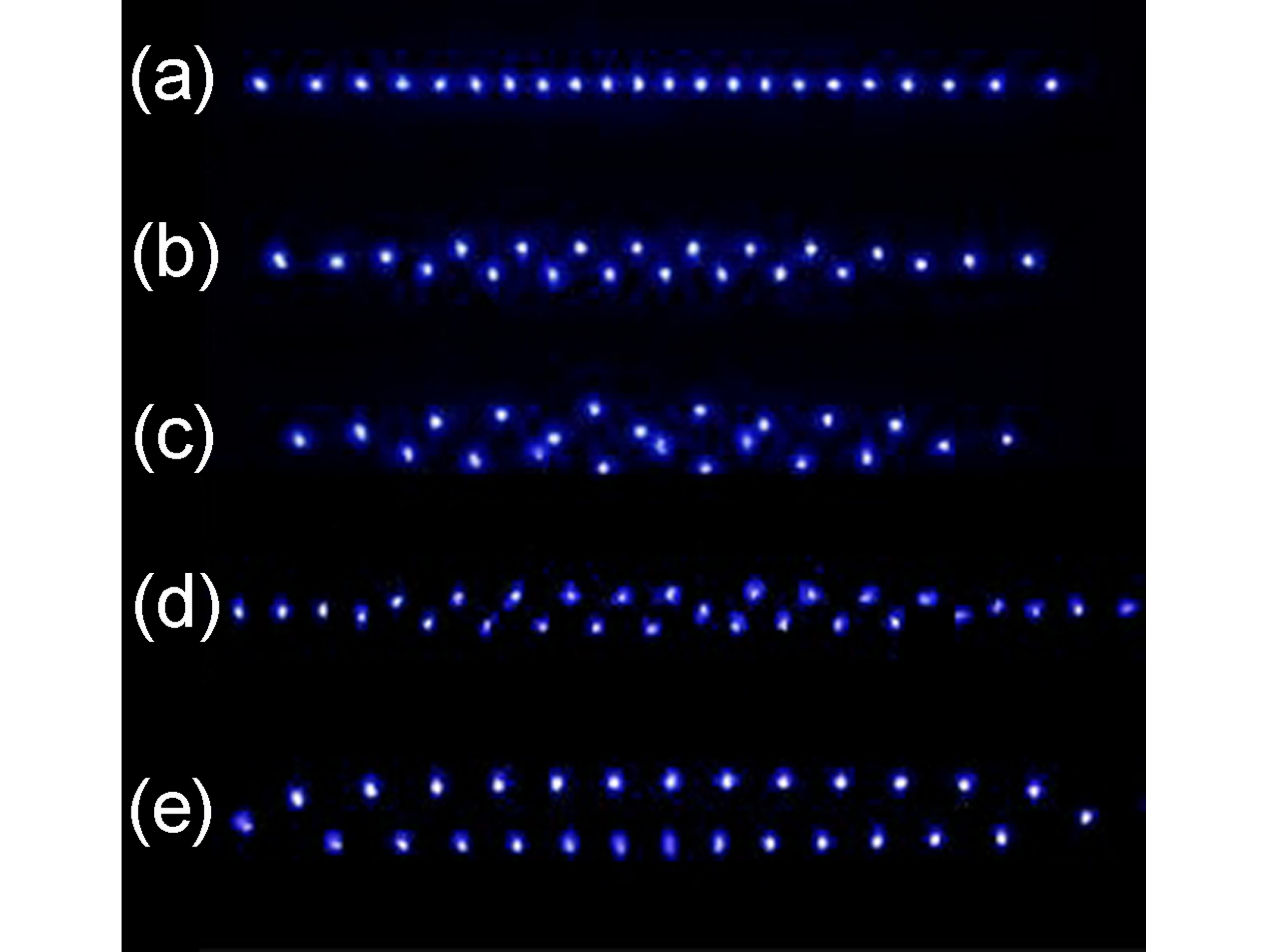}
\caption{(a) An image of a string of 22 ytterbium ions along the axis of a linear RF trap. (b) The same ions but with a slightly relaxed radial potential, for which the equilibrium configuration of the ions is a zigzag shape. (c) The ions form a helix when the potential is further relaxed.  Images (d) and (e) show defects in an ion chain, discussed in Section \ref{sec:phase} (Figure courtesy of T. Mehlst\"aubler, PTB).}
\label{fig:strings}
\end{centering}
\end{figure}

Strings of ions in a trap were first reported in 1992 \cite{waki1992,raizen1992}. A typical recent image of a string of ions in shown in Figure \ref{fig:strings}(a) \cite{pyka2013}. The ions are illuminated by laser light that is also used for laser cooling and is therefore resonantly scattered by the ions.  Very long strings can also be formed (see for example the images in \cite{schneiderreview}).  In these long strings it can be seen that the separation of adjacent ions is a smooth function of position, with its minimum value at the centre of the string.  Sometimes an ion of a different species may be incorporated into one of these strings, but does not emit any fluorescence because it  does not have a resonant response to the laser frequency. In this case the dark ion  occupies a site that would otherwise be occupied by a bright ion. The dark ion is said to be sympathetically cooled by the other (laser cooled) ions, and this happens through the Coulomb interaction, which couples the motions of all the particles \cite{drewsen15}.

The string of ions can be considered as a set of coupled oscillators and as such the small-amplitude oscillations of the system are best treated by finding the normal modes of oscillation of the string. For oscillations along the $z$ direction, the normal modes can be found starting from the total energy given in Equation \ref{eq:energy} above \cite{james}.  For all values of $N$, the first (lowest frequency) mode is the centre of mass mode at $\omega_z$ (where all the ions move together without changing their separations)  and the second mode (called the {\it breathing mode}, where the whole string expands and shrinks symmetrically with respect to the central point) is always at a frequency of $\sqrt 3\omega_z$. The remaining axial modes are at higher frequencies and involve more complicated motions.  The oscillation modes of small axial strings of ions are important for quantum information processing applications, as discussed in Section \ref{sec:QIP}.  

If the ion string is excited by an additional, weak, force (e.g. from an oscillating potential applied to  the electrodes), the string will respond resonantly when the excitation frequency matches the frequency of one of its normal modes.  By then taking images at different phases of the exciting field, it is possible to build up a sequence of images showing the evolution of the normal modes of the string \cite{nagerl1998}.  

The string also has transverse oscillations, where the motions of the ions are perpendicular to the trap axis.  These oscillations include the radial centre of mass mode, which has the single-ion radial oscillation frequency $\omega_r$.  The other radial modes are all at lower frequencies.  The lowest frequency mode, the so-called zigzag mode, has adjacent ions moving in opposite directions.  If the radial confinement is relaxed, the frequency of this zigzag mode approaches zero and at that point the ion string becomes unstable and acquires a permanent zigzag displacement \cite{retzker} (see Section \ref{sec:phase}).  This is illustrated in Figure \ref{fig:strings}(b).

\begin{figure}
\begin{centering}
 \includegraphics[width=100mm]{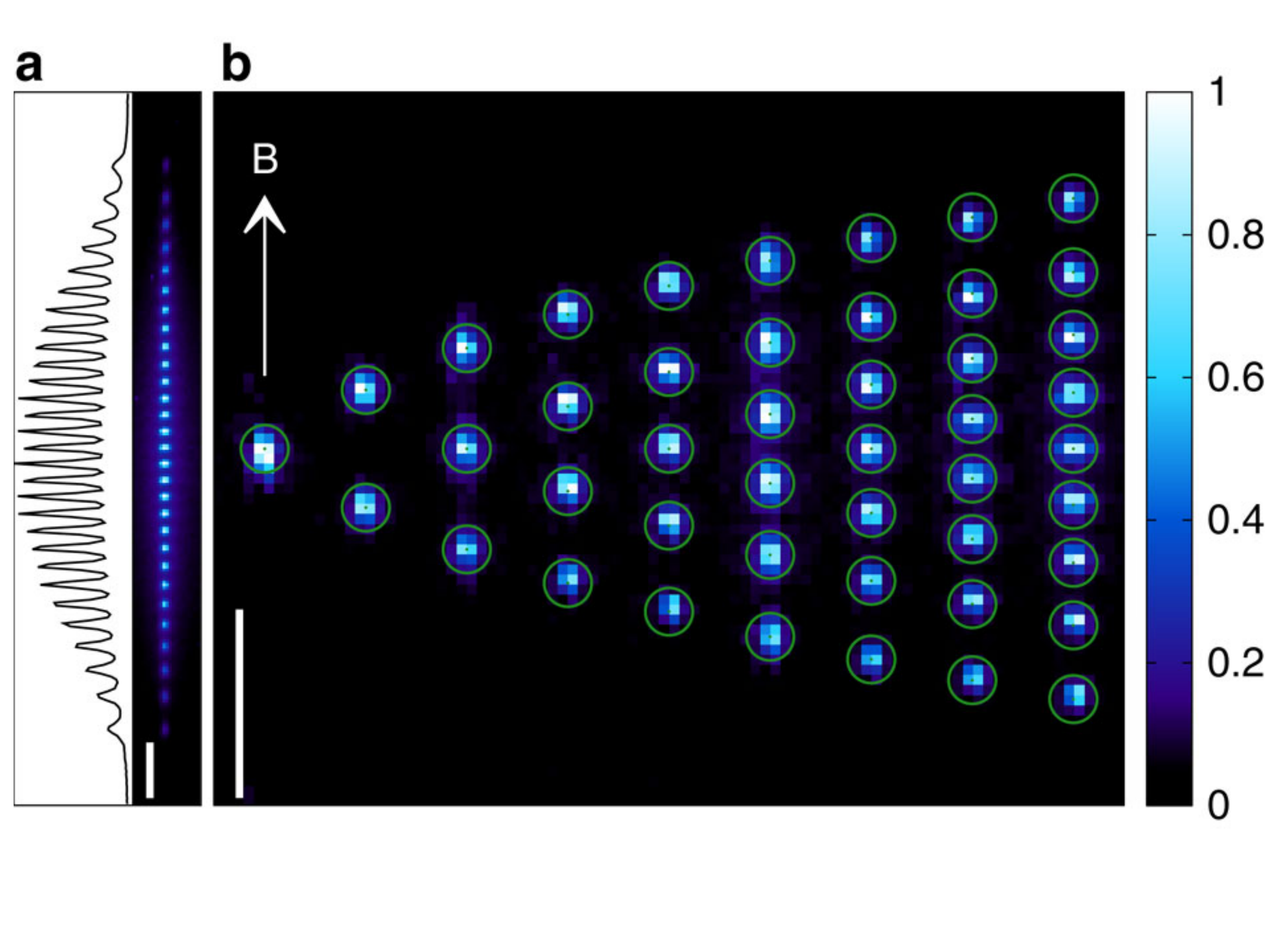}
\caption{ \textbf{(a)} Image and intensity profile of a chain of 29 ions in a Penning trap \cite{penningstrings}. The ions at the end of the chain are less bright than the central ions because of imperfections of the imaging system and because they are not  well illuminated by the radial laser beam, which has a diameter of approximately 100\,$\mu$m. \textbf{(b)} Collage of linear crystals of 1--9 ions. The applied voltage was kept constant for all of these experimental images.  Each pixel is equivalent to $2.65\pm 0.15$\,$\mu$m in the centre of the trap. The circles around the ions are the calculated positions, from Ref.\ \cite{james}, and not from a fit to the data. The bars at the bottom of the images represent a length of 50\,$\mu$m. Reprinted by permission from Macmillan Publishers Ltd: Nature Communications 4, 2571, copyright 2013.}
\label{fig:penningstrings}
\end{centering}
\end{figure}

Axial strings can also be observed in Penning traps, as has recently been demonstrated by the Imperial group \cite{penningstrings}.  Images of strings containing one to nine ions are shown in Figure \ref{fig:penningstrings}, along with a longer string containing 29 ions.  The physics of this arrangement is identical to that of the linear RF trap as the axial confinement is the same in both cases.  However, the traps differ in the way they achieve radial confinement, and this affects the formation of three-dimensional crystals (see Section \ref{sec:3DICC}).

\begin{figure}
\begin{centering}
 \includegraphics[width=120mm]{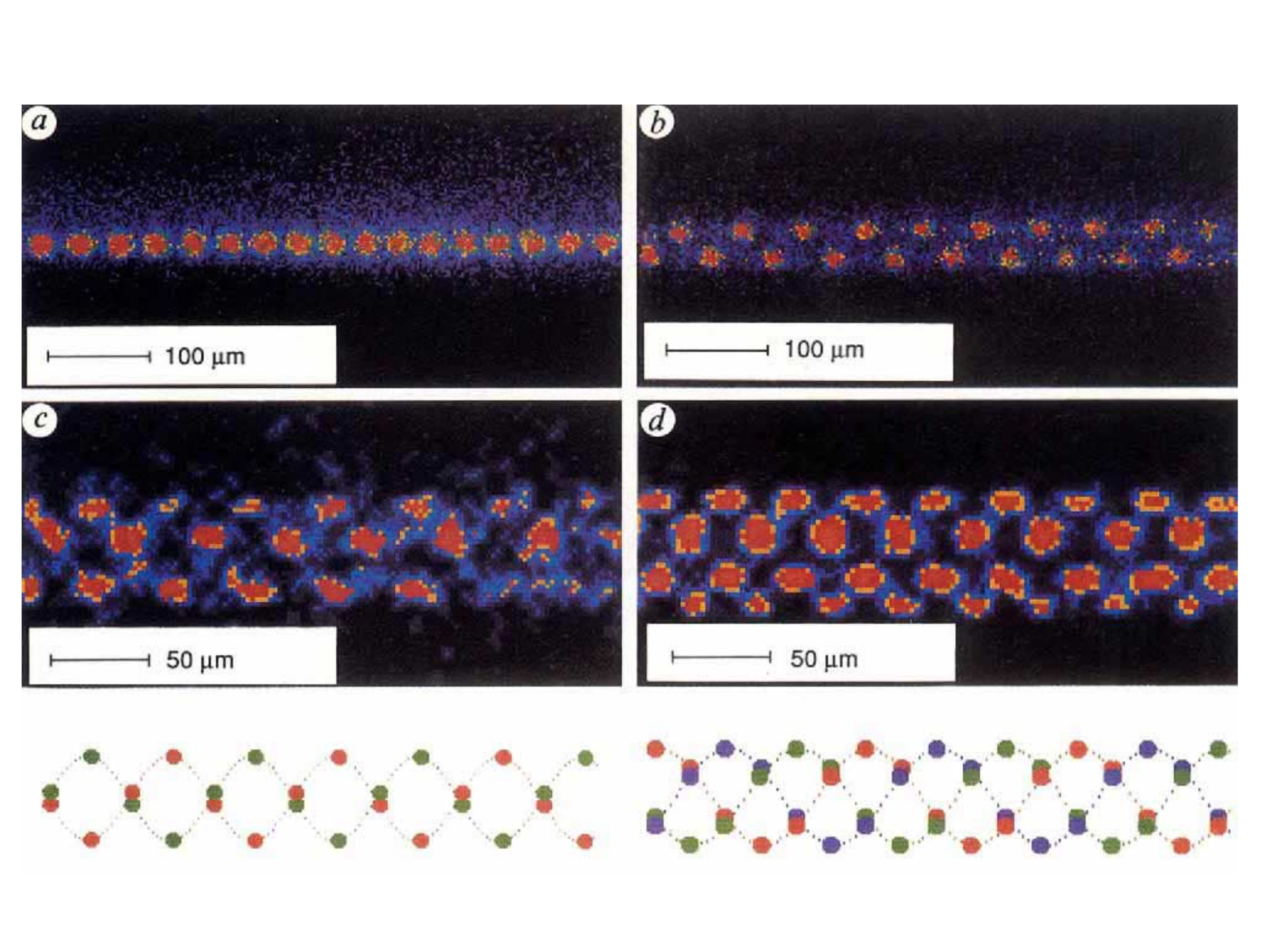}
\caption{ False colour images of long ion crystals of Mg$^+$ in a radiofrequency ring trap \cite{waltherring}. The structure of the crystal changes as the  linear density of ions increases.  In (a) the ions form a simple string; in (b) they have a zigzag configuration; in (c) the ions form a structure consisting of two interwoven helices and in (d) there are three interwoven helices.  Visualisations of the structures in (c) and (d) are given below.   Reprinted by permission from Macmillan Publishers Ltd: Nature 357, 310, copyright 1992.}
\label{fig:birkl}
\end{centering}
\end{figure}

A third type of trap in which long strings of ions can be observed is the ring trap introduced by Walther's  group at MPQ in Garching \cite{waltherring}.  This is in effect a very long linear RF trap which is curved round to make a complete ring.  It therefore does not have any axial confinement so ions are free to move around the ring.  This trap can  accommodate an extremely large number of ions in a string with a uniform spacing.  Although in principle the ions would move freely around the trap if it were perfect, in reality  there will always be places where additional potentials (e.g. from deposits on the surface of the electrodes) act to give a potential barrier that stops the ions moving past that point.  This means that the ions can be laser cooled in a similar manner to the case of the linear RF trap.  The MPQ group found that there was a critical linear density up to which the linear string was stable, and for densities greater than this the string began to kink and eventually to form cylindrical  shells, as shown in Figure \ref{fig:birkl} \cite{waltherring}.  We return to this point in Section \ref{sec:applications}.

Finally, even longer structures can be observed in  an ion beam moving in a storage ring.  This was first demonstrated using the PALLAS table-top storage ring \cite{schaetzring}, similar in design to the ring trap described above \cite{waltherring}.  The sign that crystallisation of the circulating beam had taken place was the sudden narrowing of the transverse distribution of the ions in the beam.  A comprehensive review of this topic has been given by Schramm and Habs \cite{schrammionbeams}.

\section{Three-dimensional ion Coulomb crystals}  \label{sec:3DICC}

\subsection{Single-component plasmas}

Ions confined in an ion trap constitute an example of a plasma.  However, unlike a normal plasma where there are equal numbers of positively and negatively charged particles, a cloud of ions has only one sign of charge.  In effect the presence of the confining potential is equivalent to the presence of a uniform density of particles of the opposite charge to the confined ions.  This type of plasma is therefore referred to as a \emph{single-component plasma} \cite{bollingermodes}.

It is well known from classical physics considerations that charged particles in a plasma  will form a crystal structure when cold enough.  This has been studied extensively (see, for example, \cite{crystalsims}).   The behaviour of the plasma is best described in terms of the so-called \emph{Coulomb coupling parameter}, $\Gamma$,  which is given by
\begin{equation}
\Gamma = \frac{e^2 }{4\pi\epsilon_0a_0kT}
\end{equation}
where  $a_0$ is called the \emph{Wigner-Seitz radius}, and the other symbols have their usual meanings.  The Wigner-Seitz radius is a measure of the average distance between particles and it is related to the number density $n$ by $(4/3)\pi a_0^3n=1$.  The coupling parameter is therefore essentially the ratio of the nearest-neighbour Coulomb energy to the average thermal energy of the particles.  Simulations show that plasmas behave like a gas when $\Gamma$ is below ~1 (called the \emph{weak coupling regime}) and like a liquid when $\Gamma > 2$ (called the \emph{strong coupling regime}).  At higher values of $\Gamma$ the plasma forms a regular crystal-like structure.  For an infinite plasma this phase transition occurs at $\Gamma\approx 178$ \cite{crystalinfinite}.   A simple calculation allows us to estimate the parameters required for a Coulomb crystal to form in an ion trap: assuming a typical Doppler-cooled temperature of 1~mK we find that the  inter-ion distance should be less than about 10\,$\mu$m and the density should be at least 2.5$\times 10^{14}$ m$^{-3}$  (or 2.5$\times 10^{8}$ cm$^{-3}$).  This is achievable fairly easily with linear RF traps having dimensions of the order of mm and RF drive voltages of a few hundred volt at several MHz.  

Large ICC generally have a spheroidal shape with the aspect ratio determined by the ratio of the strength of confinement (i.e. the trap oscillation frequencies) in the radial and axial directions.  They have a uniform density which depends on the trapping strength.  ICC containing a small number of ions behave in a similar manner but the exact configuration of the ions needs to be found by calculation or simulation.

\subsection{Simulations of the formation of ion Coulomb crystals} \label{sec:sims}

Extensive numerical simulations have been performed to study the formation of ion Coulomb crystals.  These have shown that  for infinite systems the minimum energy state of the crystal has a body-centred cubic crystal structure.  However, for finite systems other structures can appear such as spherical  \cite{crystalsims} or cylindrical \cite{drewsenlarge} shells.  Crystallisation occurs at a similar value of $\Gamma$  to that for the infinite case, but the process tends to start at lower values of $\Gamma$.  In these cases there may be parameters for which the central part of the cloud of ions is crystallised but not the outer part. 

The group of Schiller have carried out simulations of these systems that are able to reproduce the observed configuration of the ions in a three-dimensional crystal very accurately.  The simulations are based on a detailed calculation of the forces acting on ions, including the confining potential, the Coulomb repulsion, stochastic forces to represent heating effects such as collisions with residual gas molecules, and the interaction with the cooling laser \cite{schillersims}.  Micromotion can be included explicitly if necessary but this lengthens the time required for the simulation to run as the time step then needs to be reduced in order to follow the fast motion at the applied radiofrequency. However, its effects are not too severe when the cloud is crystallised and it can often be ignored. The simulations reproduce well the process of crystallisation of the ions and show how the crystal ends up in its final configuration.  The results of these calculations are simulated images of the ions which can then be compared to experimentally-obtained images (see Section \ref{sec:RFcrystals}).  By including the effect of heating and cooling processes, simulated images corresponding to different ion temperatures can be made.    Similar work has been carried out by other groups using different approaches to model the effect of a finite temperature \cite{schiffer2000,schuesslersims}.

\begin{figure}
\begin{centering}
 \includegraphics[width=120mm]{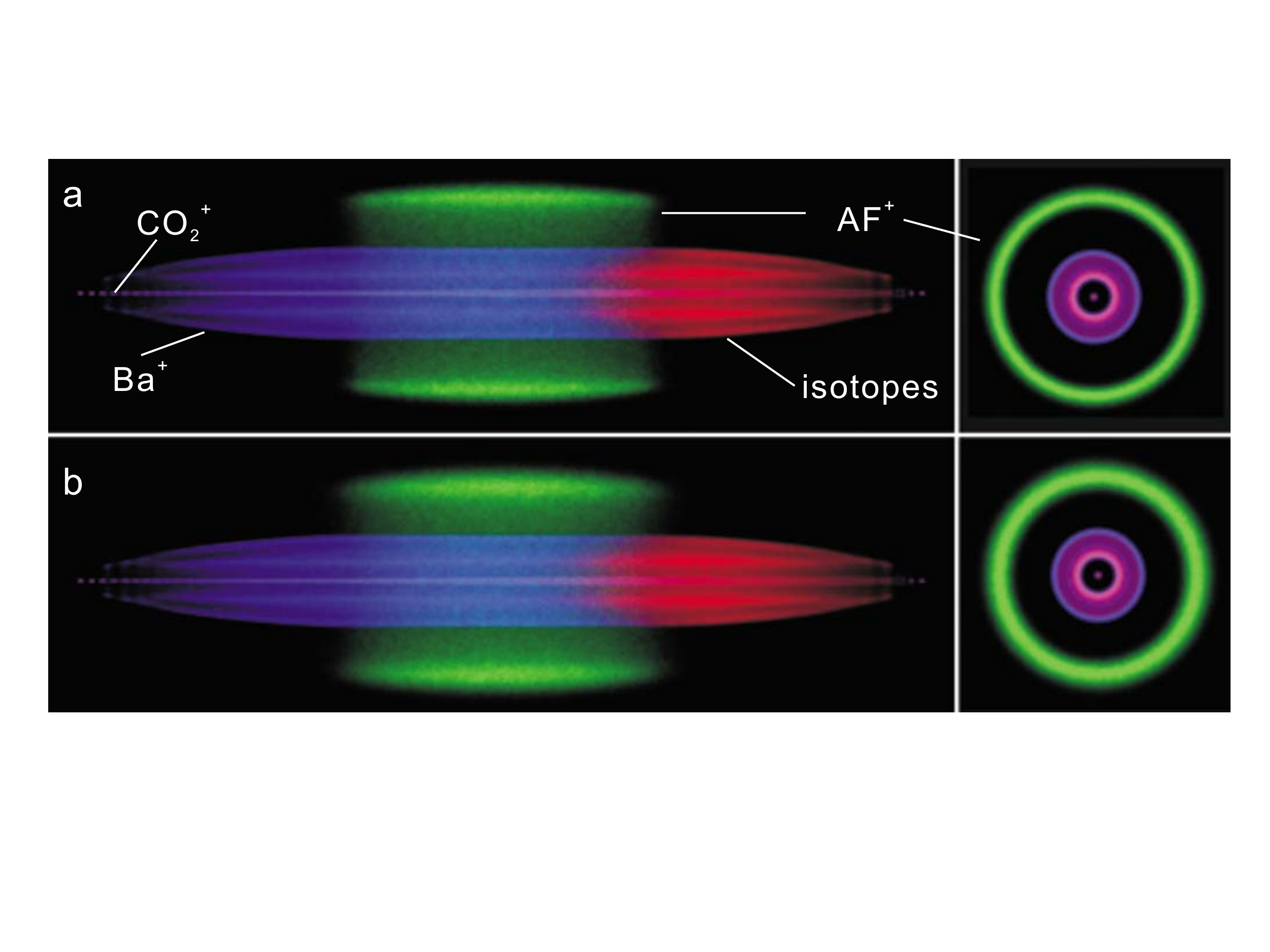}
\caption{ Simulation of an ICC containing several different species in a linear radiofrequency trap \cite{schillersims}. The crystal contains 700 laser cooled $^{138}$Ba$^+$ ions (blue),  300 sympathetically cooled  $^{137}$Ba$^+$ ions (red), 100 CO$_2^+$ ions (pink) and 200 organic molecular ions with a mass of 470 amu (green). The simulation in (b) has a higher laser cooling rate than in  (a), leading to a  higher equilibrium temperature for the sympathetically cooled species (which can be seen in the blurring of the green outer layer. Reprinted figure  with permission from C. B. Zhang \emph{et al.}, Physical Review A 76, 012719, 2007. Copyright 2007 by the American Physical Society.}
\label{fig:schiller}
\end{centering}
\end{figure}

In the simulations it is also possible to include ions of other species, in order to see their effect on the observed image, which of course does not show any fluorescence from other species as they will not be resonant with the laser. Typically, ions of different mass-to-charge ratios will form concentric shells, with the heavier ions further away from the trap axis.  The presence of a shell of high mass-to-charge ratio ions can be seen in the shape of the inner fluorescing part of the crystal, which is now flattened into a cylinder towards the centre rather than a complete spheroid \cite{schillersims}.  However, for ions with very close mass-to-charge ratios (e.g. different isotopes of the same element), this separation does not take place but the fluorescing ions are pushed towards one end of the crystal due to the light pressure from the axial laser beam.  This is illustrated in Figure \ref{fig:schiller} where different ionic species separate into different regions of the crystal \cite{schillersims}.   The behaviour of clouds of ions containing different species is an example of sympathetic cooling, which is the process by which ions of one laser-cooled species are able to cool ions of another species  in the same trap through the Coulomb interaction.

There are some differences in the way that ion Coulomb crystals form in a Penning trap compared to an RF trap, due to the fact that the whole crystal rotates.  This rotation frequency can take a range of different values and the exact configuration of the crystal (in effect, its aspect ratio) adjusts as the rotation frequency changes.  This is because the effective radial potential in the frame rotating with the crystal depends on its rotation frequency.  Simulations of crystals in a Penning trap must take this rotation into account.  In Ref \cite{penningstrings} the many different structures observed for  a 15-ion crystal could be reproduced in simulations, and this allowed the rotation frequency of the crystal to be estimated (see Section \ref{sec:penningcrystals}).  In this case the simulations do not attempt to follow the process of crystallisation, but simply find iteratively the lowest energy configuration of the system under the influence of the trapping fields, the crystal rotation and the Coulomb repulsion.

Other physical systems can also show this sort of crystallisation.  The idea was first suggested by Wigner in the 1930s \cite{wigner1934,wigner1938} in a discussion about the states of electrons in metals.  Realisation of a Wigner crystal for electrons in three dimensions is difficult due to quantum mechanical effects.  However, two-dimensional crystallisation of electrons on the surface of a superfluid has been observed \cite{wigner2dobs}. For a full discussion about the theory and a review of experimental evidence for Wigner crystallisation of electrons, see the review by Tsidil'kovski{\u \i} \cite{wignerreview}.

\subsection{Observations of crystals in radiofrequency traps} \label{sec:RFcrystals}

Many groups have observed and studied the formation of ion Coulomb crystals in radiofrequency ion traps.  The first observations were made in a three-dimensional Paul trap, where the strength of the confinement is similar in all three dimensions and the micromotion is zero only at the exact centre of the trap rather than along a line as in the linear RF trap.  Images of small numbers of laser-cooled ions in different configurations  were obtained in this way \cite{wineland1987,diedrich1987}. 

Much larger numbers of ions can be crystallised in linear radiofrequency ion traps due to the lower level of micromotion in these traps.  Images have been obtained for more than 10$^5$ ions in these traps.  Given the linear symmetry of the trap itself, the ions tend to arrange themselves in open or closed cylindrical shells \cite{drewsenlarge}.  Since the trapping potential felt by an ion depends on the mass to charge ratio for that ion, different species experience different forces and this results in a separation of the different species, as discussed in Section \ref{sec:sims}.  It is notable that even though laser cooling can only be applied to one species, a cloud of ions consisting of more than one species can be maintained at a temperature that is low enough for the whole system to remain crystallised by means of sympathetic cooling.  For example, in one experiment a cloud of 15 ions was kept crystallised by a single laser-cooled ion in this way \cite{drewsen15}.

\begin{figure}
\begin{center}
\includegraphics[width=120mm]{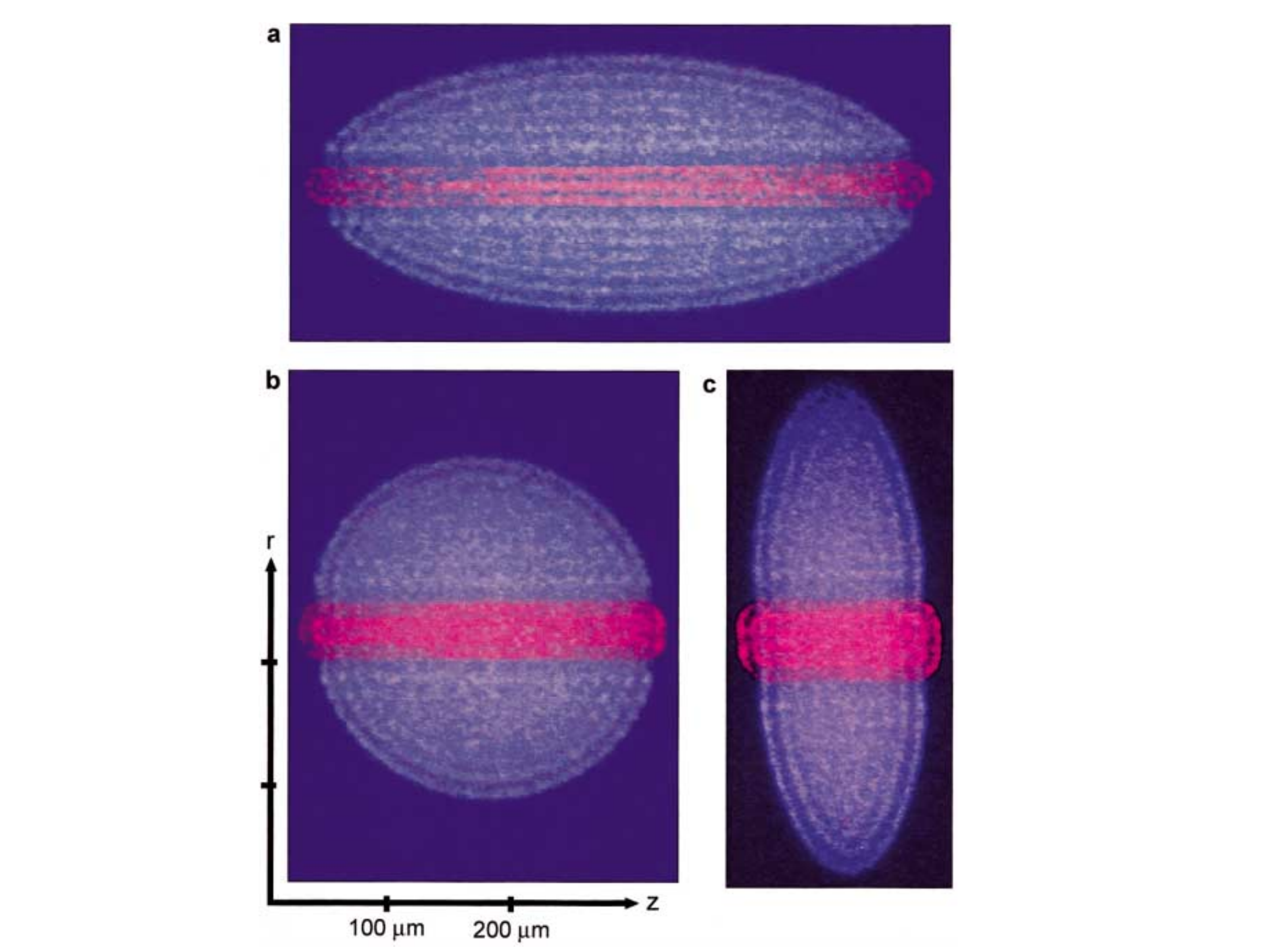}
\caption{A $^{40}$Ca$^{+}$/$^{24}$Mg$^{+}$ bi-crystal at three different end cap voltages in a linear ion trap \cite{drewsenlarge}.  The crystal is symmetric under rotations about the trap axis, $z$, and contains approximately 300 $^{24}$Mg$^{+}$ ions (red) and 3000 $^{40}$Ca$^{+}$ ions (blue).  The aspect ratios of the crystals are determined by the ratio of axial to radial trapping frequencies for the two species.  This parameter is always higher for $^{40}$Ca$^{+}$ than for $^{24}$Mg$^{+}$ and it increases (for both species) in the sequence (a) to (c). Reprinted figure  with permission from M. Drewsen \emph{et al.}, Physical Review Letters 81, 2878, 1998. Copyright 1998 by the American Physical Society.}
\label{fig:drewsen}
\end{center}
\end{figure}

Figure \ref{fig:drewsen} shows images of an ion crystal containing two species ($^{40}$Ca$^{+}$, blue, and $^{24}$Mg$^{+}$, red).  The images are obtained by superimposing light from the two different species in two separate exposures.  The strength of the confinement for Mg$^+$ is always stronger than for Ca$^+$ due to its lighter mass, so it is always positioned  along the trap axis with the Ca$^+$ around it in a ring.  Note that the overall shape of the crystal is always roughly ellipsoidal even though it consists of two different species \cite{drewsenlarge}.

In the experiments of the Schiller group in Du\"sseldorf it was shown that is  possible to determine the exact number of ions present by careful observation of the detailed shape of the crystal and comparison with the results of computer simulations \cite{schillersims}.  In the same manner, comparison of simulated images of ion crystals that include thermal effects with experimentally obtained images allows the ion crystal temperature to be estimated. Because the number of ions in the trap can be determined so accurately, this system can be used to study the rates of chemical reactions between trapped ions and neutral gas in the chamber (see Section \ref{sec:reactions}).

\subsection{Observation of crystals in the Penning trap} \label{sec:penningcrystals}

  Crystals of ions in linear RF traps tend to be prolate, that is, long and thin like a cigar.  This is because the strength of the radial confinement is generally much greater than that of the axial confinement.  In the extreme case, the crystal consists of a long string of ions on the axis (see Section \ref{sec:1DICC}). On the other hand, the relative strengths of the axial and radial confinement in a Penning trap can be adjusted over a wider range, and in particular it is possible to adjust the parameters of the trap to change the configuration of a small crystal from a linear string, through different three-dimensional structures, to a planar crystal.   The Bollinger group at NIST has been most active in this area and has worked routinely with crystals that consist of a single two-dimensional plane containing hundreds of ions \cite{bollingerdrumskin}.  

There are two main problems with the creation and observation of ion Coulomb crystals in the Penning trap.  The first problem is that laser cooling is not very effective for the magnetron motion, which for a single ion is a slow orbit around the centre of the trap.  Because the total energy of this motion is negative, energy has to be supplied to the ion to reduce the amplitude of the magnetron motion.  One way to do this is to offset the laser cooling beam radially from the centre of the trap \cite{itano1982,papa}. This has the effect of applying a torque to the ion as well as providing damping.  For large clouds of ions the equivalent of the magnetron motion is a global rotation of the whole cloud, which now behaves like a plasma.  Note that in the Penning trap the crystal always rotates, due to the presence of the magnetic field. This rotation frequency, $\omega_r$, is linked to the number density, $n$, through the relation \cite{bollingermodes}
\begin{equation}  \label{eq:density}
n=2\epsilon_0m\omega_r(\omega_c-\omega_r)/e^2
\end{equation}
The density of the cloud is therefore maximum when the rotation frequency is $\omega_c/2$, and this condition is termed \emph{Brillouin flow}.  It is often possible to reach this condition using just a radially offset laser beam.

In the case of large clouds it is  more effective to apply the torque in a different way, and the NIST group has developed the \emph{rotating wall technique} for this purpose \cite{bollingerrotatingwall}.   Here a radial linear or quadrupole electric field, rotating at some frequency $\omega_{RW}$, perturbs the ion plasma and forces it to rotate at the same frequency.  Since this rotation frequency is linked to the density of the ions through Equation \ref{eq:density}, it has the effect of controlling the ion density and also the shape of the plasma \cite{bollingerrotatingwall,bharadia2012}. Laser cooling can then be provided with an axial beam; if the cooling is strong enough, the plasma will  crystallise in the same way as in an RF trap.   The rotating wall therefore allows the density and shape of the ion Coulomb crystal to be controlled.

\begin{figure}
\begin{centering}
 \includegraphics[width=100mm]{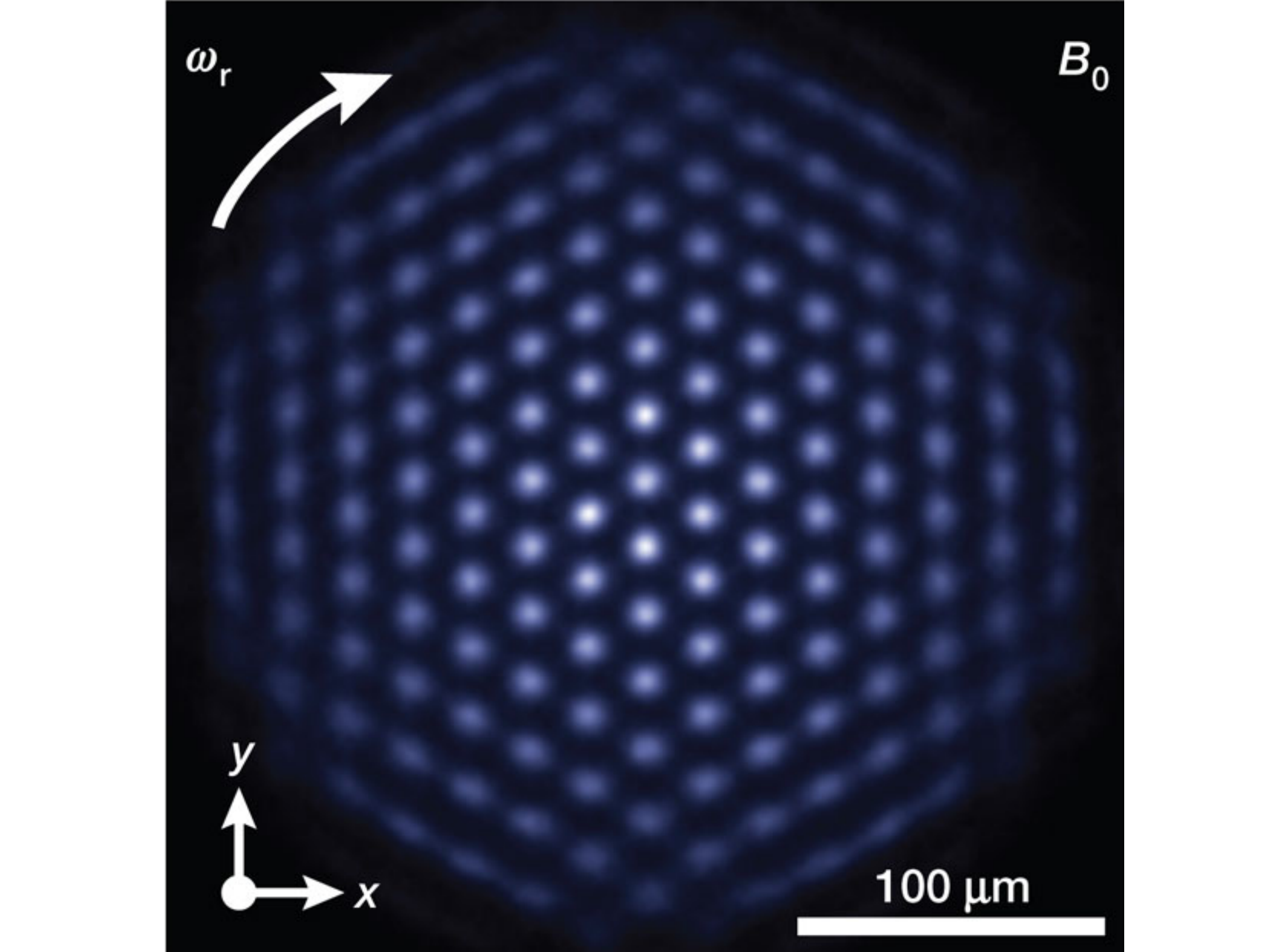}
\caption{ An image of an ion  Coulomb crystal consisting of a single plane of Be$^+$ ions  in a Penning trap \cite{bollingersims}.  The crystal is rotating but the image is reconstructed using timing information from a position-sensitive photomultiplier. Reprinted by permission from Macmillan Publishers Ltd: Nature 484, 489, copyright 2012.}
\label{fig:hexagonal}
\end{centering}
\end{figure}

The second problem with the Penning trap is also a consequence of the fact  that the crystal is always rotating.   The rotation means that the crystal cannot easily be imaged without blurring, so in order to obtain clear images it becomes necessary to gate the camera so that the ions are always imaged at the same point in their rotation.  Since the rotation is locked to the applied rotating wall frequency  ($\omega_{RW}$), this is relatively straightforward provided the camera is capable of fast gating. As an alternative, a position-sensitive photomultiplier can be used for imaging.  In this way the Bollinger group have obtained spectacular images of rotating crystals consisting of a single plane of ions (see Figure \ref{fig:hexagonal}) \cite{bollingersims}.  A strong advantage of the Penning trap for this type of work is the absence of micromotion, which would be significant in an RF trap for the ions far removed from the axis of the trap. 

Ion crystals in the Penning trap can form in a number of different configurations.  In particular, for a two-dimensional crystal consisting of a single layer of ions, a triangular  lattice is normally formed.  As the rotation frequency of the crystal ($\omega_r$) is increased (thus effectively increasing the strength of the radial confinement), the crystal splits into two planes, but also changes to a square lattice \cite{bollingercrystal}.  As $\omega_r$ is further increased, the square lattice becomes rhombic and eventually triangular again before undergoing another split into three planes.  This process then repeats.

The large single-layer planar crystals have interesting transverse modes of vibration and behave rather like vibrating drum-skins \cite{bollingerdrumskin}.  The frequencies of these modes can be calculated and compared to experimentally measured values.  In their experiments, the Bollinger group at NIST worked with a planar crystal containing several hundred ions in a single layer.  With an inter-ion spacing of around 20\,$\mu$m, the diameter of the crystal is a few hundred $\mu$m.  The modes were excited using an optical dipole force from a pair of laser beams inclined at a small angle to the plane of the crystal and detuned from the optical resonance of the Be$^+$ ions at 313\,nm.  The dipole force is spin-state dependent so by initially placing all ions in a superposition of their spin states, any mode can be excited by choosing a suitable frequency difference between the two beams.  In this way the effective temperature of each mode could be determined.  The measured temperatures were consistent with the expected Doppler cooling limit of 0.4\,mK, except for the centre of mass mode, which had a higher temperature of a few mK.  This arises because the pulses of light used for laser cooling were turned on and off in a very short time.   The resulting radiation pressure, which was uniform across the whole crystal, excited the centre of mass mode, increasing its temperature above the Doppler limit. In recent work the NIST group has carried out quantum simulation experiments using this system (see Section \ref{sec:QIP}).

Conventional crystals of atoms can be investigated by means of Bragg scattering of x-rays.  This is a well-established technique for determining crystal structures: from the angles of scattered rays information can be obtained on the spacing of the atoms and the type of crystal lattice. In order for the technique to be effective, the wavelength of the radiation used must be comparable to the typical distances between atoms in the crystal (i.e. less than 1~nm) -- hence the use of x-rays.  Bragg scattering can also be used to study ion Coulomb crystals and in this case, since the typical distance between ions is of the order of a few $\mu$m, optical radiation has a suitable wavelength, giving scattering angles of a few degrees.  This means that radiation at the ion's resonance wavelength, which is used for laser cooling, can also act as the incident radiation for Bragg scattering.  

The NIST group has used this technique with a three-dimensional crystal of Be$^+$ ions in a Penning trap, to demonstrate different configurations of the crystal.  Bragg scattering from ions in RF traps is not possible due to the RF micromotion and the lack of long-range order in the crystals.  In Penning traps,  medium-sized crystals with  roughly spherical symmetry tend to form in concentric shells.  Long-range order starts to set in at a diameter of around 37$a_0$ but the predicted body-centred cubic (BCC) bulk configuration does not form unless the crystal has a diameter of around 65$a_0$, meaning that the crystal contains at least 300~000 ions.  There is clear evidence that under these circumstances the configuration is a BCC crystal \cite{bollingerbragg}.  This can be deduced from the Bragg scattering pattern, which is a series of concentric rings (due to the rotation of the crystal) with characteristic angular spacings.  

For smaller numbers of ions, a number of other crystal configurations were observed, because in this case the presence of the surface affects the arrangement of ions.  These included crystals with 5-fold symmetry, observed through both the Bragg pattern and direct imaging of the crystal \cite{bollingercrystal}.

\begin{figure}
\begin{centering}
 \includegraphics[width=170mm]{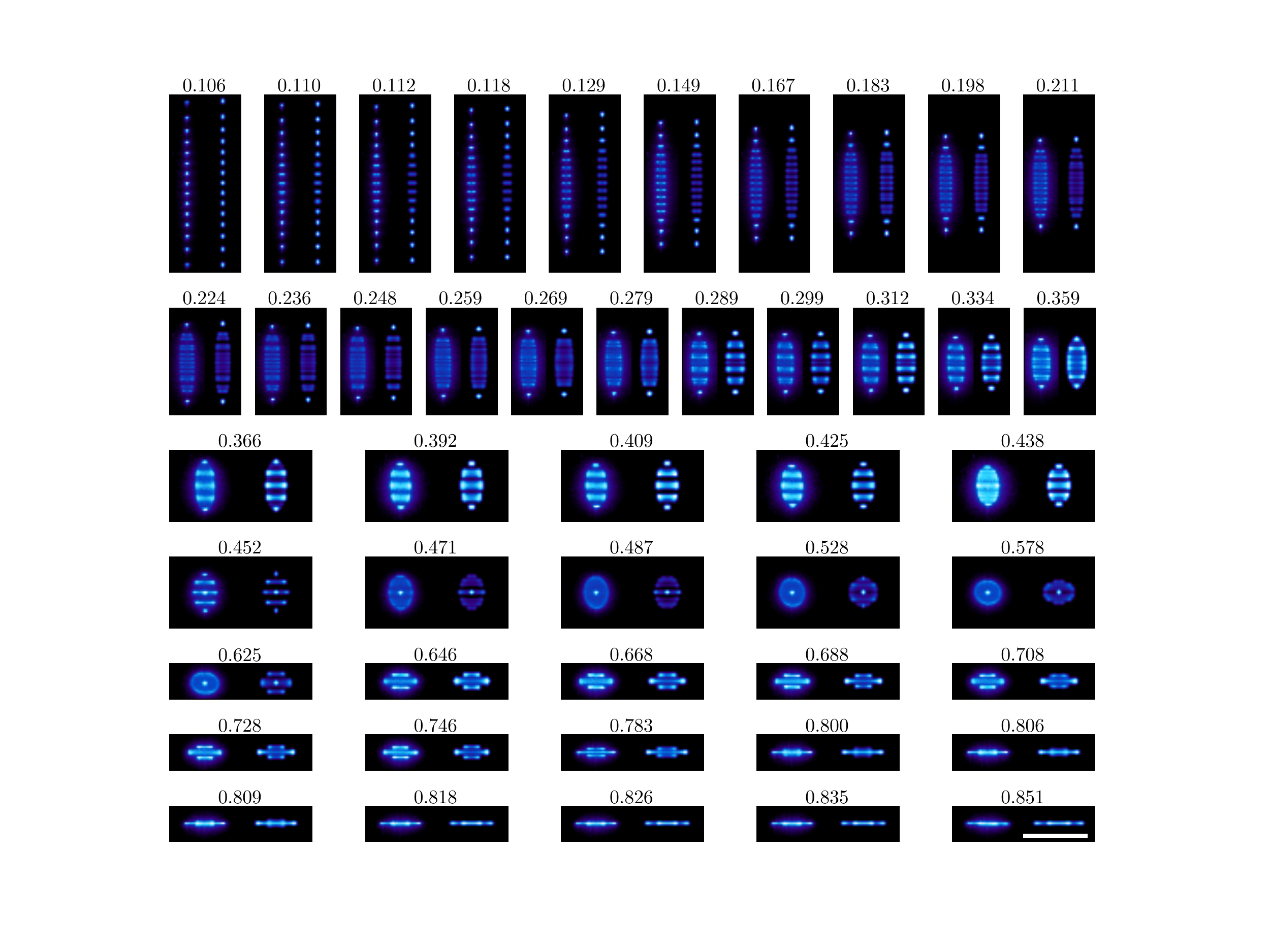}
\caption{ Conformations of a 15-ion crystal in a Penning trap \cite{penningstrings}.  Experimentally obtained images  (left side of each pane)  are compared to computer simulations (right side of each pane). By increasing the axial confinement a linear string is transformed into a zigzag structure, then a 3-D crystal and finally a planar structure. Each image is labelled with the value of the normalised axial trapping frequency, which increases with the trapping voltage (the trap becomes unstable when this quantity is equal to unity). There is a 100\,$\mu$m scale bar in the bottom right-hand pane which applies to all the images.  Reprinted by permission from Macmillan Publishers Ltd: Nature Communications 4, 2571, copyright 2013.}
\label{fig:zoo}
\end{centering}
\end{figure}

At the other extreme, the Imperial group has used a Penning trap to prepare and image small numbers of ions (up to 20 ions) in different configurations as the trapping parameters are varied.  For low axial potential the expected configuration is an axial string, as explained above.  As the axial potential is increased, first the string of ions kinks into a zigzag shape, and then it forms several three-dimensional structures (depending sensitively on the number of ions) until eventually it becomes a planar crystal.  These different configurations can also be obtained from simulations that attempt to find the lowest energy configuration of the crystal for given trapping parameters, and a good agreement between the simulations and observations has been obtained (see Figure \ref{fig:zoo}).    In these experiments the rotation frequency of the ion crystal in the laboratory frame is not known, because there is no rotating wall applied.  It can take a range of values, depending on the size, position and frequency of the laser beam, and this can be checked for consistency with the observed structures \cite{penningstrings,asprusten2014}.  Note that for a small number of ions it is not strictly appropriate to describe the crystal shape as a spheroid.  However, the overall aspect ratio and density of the ions are always very  close to what would be expected for a larger number of ions, once surface effects are accounted for.

\subsection{Coulomb crystals of macroscopic particles}

\begin{figure}
\begin{centering}
 \includegraphics[width=100mm]{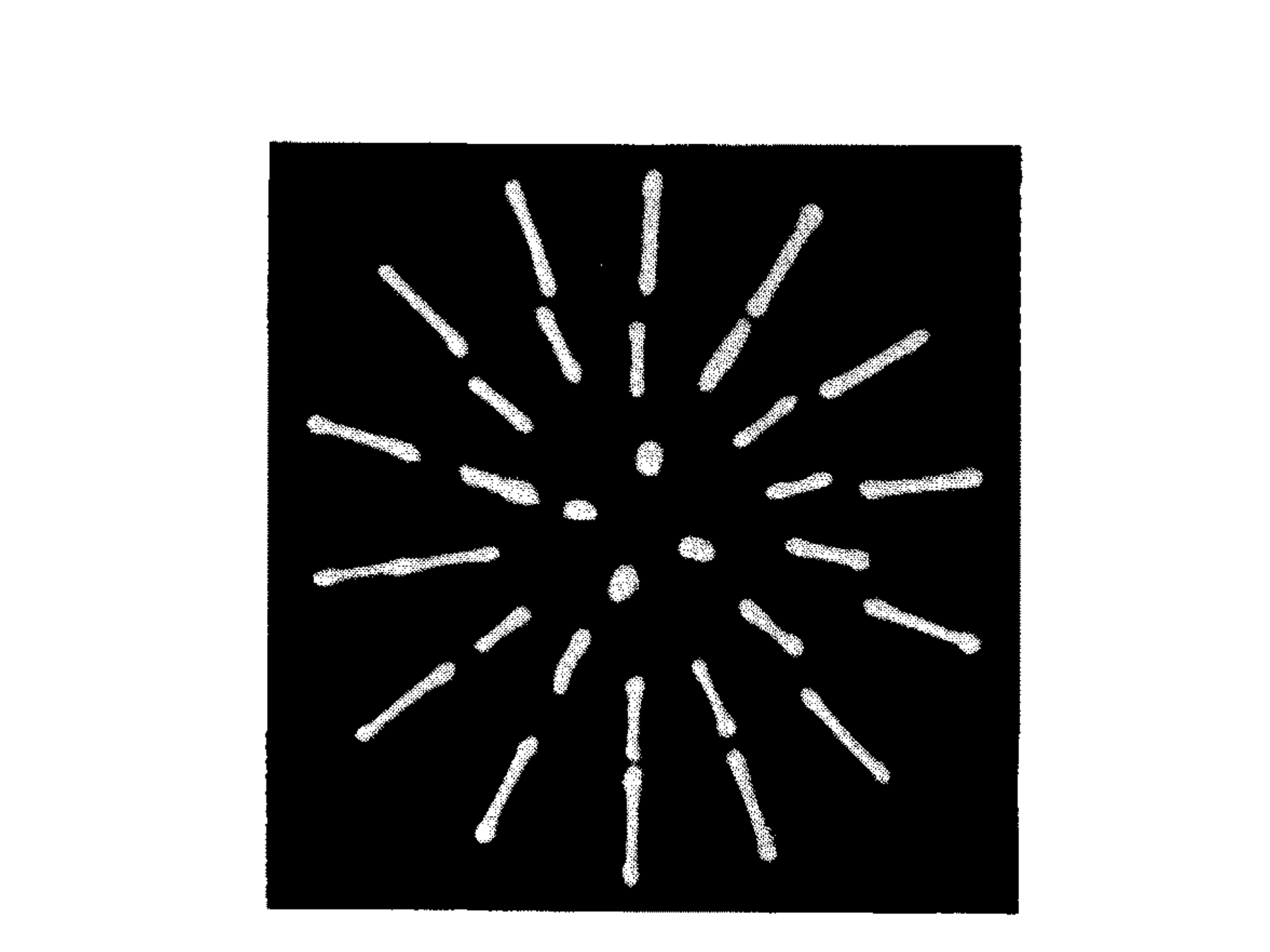}
\caption{ Photograph of 32 charged aluminium particles in a radiofrequency trap, viewed in the radial plane \cite{wuerker1959}.  The micromotion of each particle results in its image spreading into a line. Reprinted with permission from Journal of Applied Physics 30, 342. Copyright 1959, AIP Publishing LLC.}
\label{fig:wuerker}
\end{centering}
\end{figure}

There is nothing in the equations of motion of particles in RF traps that restricts the nature of the trapped particles to be atomic ions.  At around the same time that Paul was developing the first three-dimensional trap for atomic ions, a different research group was also working on a trap for charged dust particles with dimensions of the order of micrometres \cite{wuerker1959}.  The physics of this trap is identical to that of the now familiar RF trap but the parameters are rather different:  because the mass to charge ratio of the particles is much higher, the frequency of the applied potential is typically around 100\,Hz and the required amplitude is a few hundred volts.  Damping is provided by the air.  Figure \ref{fig:wuerker} shows a photograph  from this paper, demonstrating that the structures formed are very similar to those formed by atomic ions.  Note that in this photograph each particle's image becomes a line because of the relatively large micromotion in this system.

\section{Applications of ion Coulomb crystals}  \label{sec:applications}

\subsection{Phase transitions in ion Coulomb crystals}  \label{sec:phase}

As has been mentioned above, ion Coulomb crystals undergo what might loosely be called phase transitions between different configurations as the trapping parameters are changed.  Strictly, this term should only be used in a system containing large numbers of particles but it is also convenient to use it here.  Consider first a long string of ions in a linear RF trap (as in Figure \ref{fig:strings}(a)).  As the strength of the axial confinement of the ions is increased, there is a well-defined point at which the string changes into a zigzag shape (see Figure \ref{fig:strings}(b)).  This would typically be referred to as a second-order phase transition.  Initially the degree of kinking into the second dimension is very small but rises rapidly as the axial confinement is further strengthened.  In the zigzag configuration (provided that the radial potential does not have perfect radial symmetry), the string has two stable states which are mirror images of each other; these two states are degenerate in energy.  Further, more complicated, phase transitions follow as the crystal becomes more of a solid three-dimensional structure  (as shown in Figure \ref{fig:strings}(c) where the ions form a helical structure).   Figure \ref{fig:birkl} shows a similar process of evolution between different crystal structures for large numbers of ions in a radiofrequency ring trap.

This linear to zigzag transition and the phase diagram of the system have been extensively studied theoretically (see, for example, \cite{morigi2004,morigi2008}).  Unfortunately the spacing of the ions is not uniform throughout the string, but it is possible to make the approximation that close to the centre of the string the linear density of ions is roughly constant, and this allows useful calculations to be performed.  Recently, these classical calculations of ion string dynamics have been extended into the quantum regime (see, for example, \cite{morigi2011}).

The structural transition between the linear string  and the zigzag structure can be used for a completely different sort of study,  illustrated in Figure \ref{fig:strings}(d) and (e).   If the confining potential of a linear string is \emph{rapidly} changed across the transition point, different regions of the chain will evolve into a zigzag at the same time, but will not be able to communicate if the quench is fast enough.  Therefore there will be defects in the crystal where regions with opposite zigzag displacements meet.  Figure \ref{fig:strings}(d) shows a localised defect where two such regions meet (referred to as an \emph{odd defect}), and Figure \ref{fig:strings}(e) shows an \emph{extended defect} where the orientation of the zigzag changes slowly.  The formation of these defects is closely related to the formation of defects in the early universe  through the process of spontaneous symmetry breaking (the so-called Kibble-Zurek mechanism; see \cite{kibble1976}).  The density of defects created depends on the speed of the quench from the linear to zigzag configurations.  Recent experimental studies \cite{pyka2013, ulm2013} have shown that this system reproduces the features expected theoretically for  the Kibble-Zurek mechanism applied to ion chains \cite{morigikz} and have confirmed the calculated scaling with quench time.

The simplest possible example of this system is a string of three ions, and this is amenable to calculations \cite{retzker}.  In a potential corresponding to an axial frequency $\omega_z$ and with a transverse frequency $\omega_x$, it can be shown that the critical point, where the crystal kinks, is given by $\omega_x^2/\omega_z^2$ =2.4 \cite{rafac}.  If one considers the potential energy of the whole system, it can be seen that at low values of $\omega_z$ there is a potential well corresponding to transverse vibrations of the string in a \emph{kink mode} (where the central ion moves in the opposite direction to the other two ions).  This is the so-called \emph{soft mode} and is the lowest frequency of vibration of the system.  As the critical point is approached, this vibration frequency drops and the potential becomes flatter.  At the critical point, it is no longer quadratic at the centre, but  quartic.  Beyond this point, the potential has a double well and this means that there are two degenerate configurations, which are mirror images of each other and have the central ion displaced from the trap axis in the opposite direction to the other two ions.  This  offers the intriguing possibility to observe quantum mechanical tunnelling between these two degenerate configurations of the three-ion system.  However, this is technically very difficult as the spacing of the double well potential has to be comparable to the spread of the ionic wavefunction in order for the tunnelling rate to be observable \cite{retzker}.

\subsection{Cavity quantum electrodynamics} \label{sec:CQED}

\begin{figure}
\begin{centering}
 \includegraphics[width=80mm]{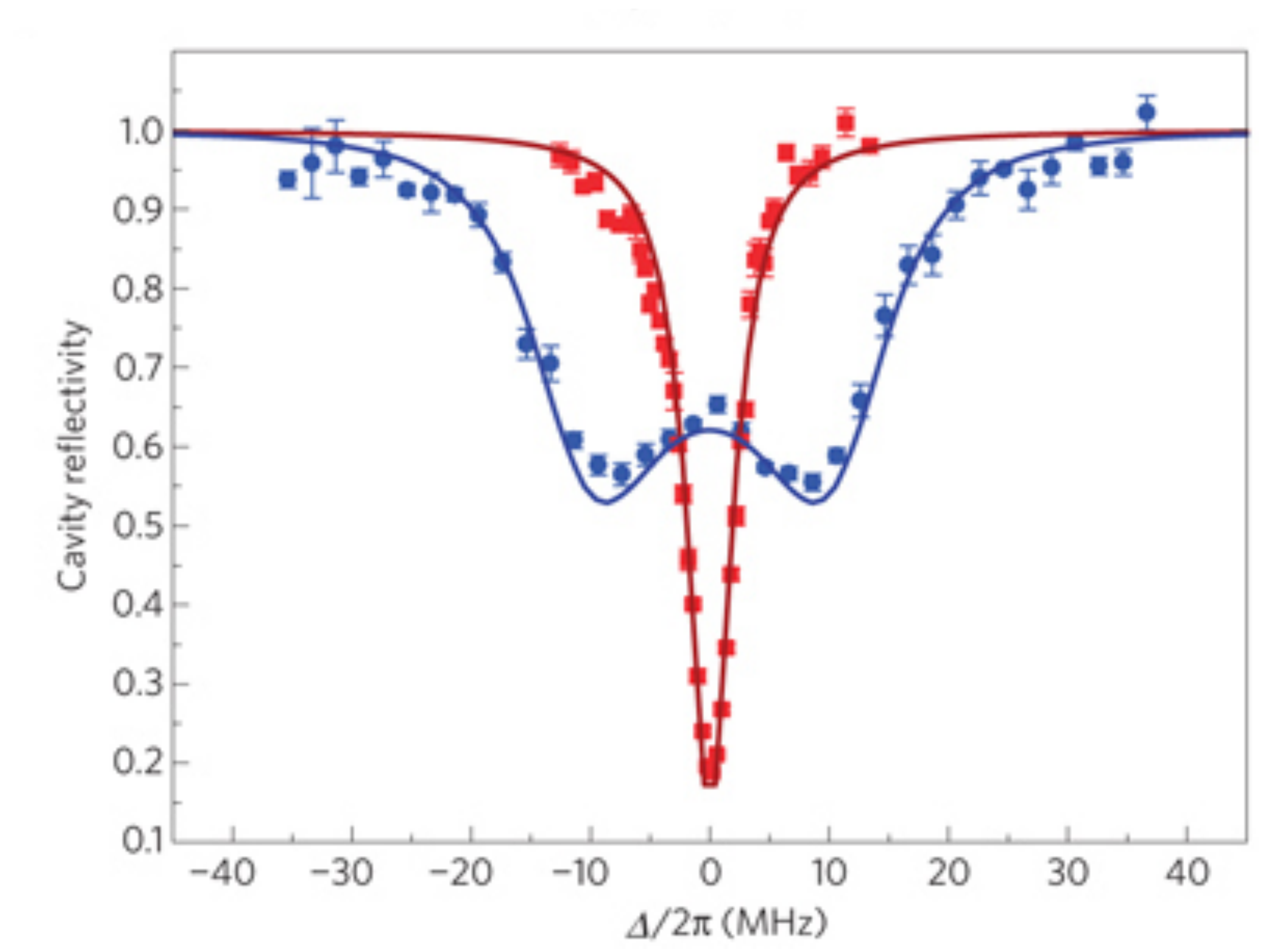}
\caption{Reflectivity of an optical cavity both without (red) and with (blue) an ion Coulomb crystal present in the cavity \cite{drewsenCQED}.  The cavity mode is in resonance with the transition in the ion and the frequency of the probe light (containing a maximum of one photon in the cavity) is tuned around the resonance.  The change from a single dip (consistent with the cavity finesse of roughly 3000) to a double dip is indicative of the system reaching the collective strong coupling regime, where the exchange of energy between the cavity and the crystal is faster than the decay of the cavity.  Reprinted by permission from Macmillan Publishers Ltd: Nature Physics 5, 494, copyright 2009.}
\label{fig:cavity}
\end{centering}
\end{figure}

Ions in a Coulomb crystal constitute a system with unusually attractive properties for many experiments:  in particular, they are nearly stationary, they are well isolated from each other and from the environment, and their internal electronic states can be manipulated with a high degree of control.  This makes them ideal for use in experiments in the area of cavity quantum electrodynamics (CQED), which deals with the interaction between atoms and light in an optical cavity, at the level of single of single photons.  The first experiment in this area, carried out by Drewsen's group at Aarhus, demonstrated that by incorporating a high-finesse optical cavity into an ion trap structure, ions could be located precisely along the axis of the cavity so they interact strongly with the cavity light \cite{drewsenCQED}.  Since the ions act cooperatively, it was possible to reach the collective strong coupling regime, where the coherent exchange of energy  between the ions and the optical cavity was faster than the decay of light in the cavity.  For this, a crystal containing at least 500 ions was required.  At any time, the cavity contains at most one photon.  Figure \ref{fig:cavity} shows the change in the cavity reflectivity around a cavity resonance with and without the presence of ions. This plot is a clear demonstration that the strong coupling regime has been reached.

This system has a number of applications apart from a demonstration of the physics of CQED, including a technique for studying the vibrational modes of large Coulomb crystals in a sensitive and non-invasive manner \cite{dantan2010}.  It has also been used to observe electromagnetically induced transparency (EIT), where a separate optical control field can switch the atomic absorption on and off in a controlled manner \cite{albert2011}.

\subsection{Quantum information processing}  \label{sec:QIP}

A classical computer uses an array of two-state systems (classical \emph{bits}) to carry out calculations using binary arithmetic.  A quantum computer is a device that uses a system of simple two-state quantum mechanical systems (e.g.  spin--1/2 particles) as \emph{qubits} for a similar purpose.  There are several reasons why the construction of a quantum computer would be very desirable, and one of these is that some computational problems are effectively impossible for a classical computer (e.g. the factorisation of very large numbers that are the product of two prime numbers), whereas they are believed to be tractable for a modest quantum computer using specially designed algorithms that take advantage of the way that quantum mechanical systems behave.  Just like classical computers work using classical gate operations (such as AND, OR and NOT) on classical bits, so quantum computers use quantum gates acting on qubits \cite{CPozeri}.  Quantum computing (or more generally \emph{quantum information processing}) has now become a huge field of study, with various physical systems being investigated as realisations of qubits (see for example \cite{ladd2010}), but we do not attempt to review it in detail here.

Many demonstrations of the essential operations necessary for building a quantum computer have been carried out using trapped ions \cite{monroe2013}.  This field of application of trapped ions was initiated by Cirac and Zoller in 1995 \cite{ciraczoller}.  They proposed the realisation of a CNOT gate, one of the basic elements of a quantum computer, using two ions in an RF trap. The internal electronic state of each ion forms  one qubit, with the ion's ground state representing a logical 0 and a long-lived excited state (an excited electronic, hyperfine or Zeeman state) representing the logical 1.    An essential component of their proposal was the use of the common vibrational mode of the two-ion system as a means of communicating between the ion qubits.  It was therefore necessary to be able to cool the system to its lowest vibrational quantum state, for which optical sideband cooling would be essential.  The first direct realisation of the Cirac-Zoller scheme was carried out by Blatt's group at Innsbruck \cite{blattCZ}.  Since then, ion Coulomb crystals have been used in realisations  of increasingly complex demonstrations of quantum computing algorithms.  For a detailed review of this aspect of the use of ICC,  see the reviews by Roos \cite{roosPTCP} and Ozeri \cite{CPozeri}.  

One of the requirements for a practical quantum computer is for it to be scalable to large numbers of qubits, which in this case means large numbers of ions.  However, as the number of ions in a string rises, the number of possible modes of oscillation also increases.  Since each mode has a different frequency, it is difficult to separate out individual sidebands, which makes sideband cooling more difficult.  This puts a practical limit on the number of qubits that can be manipulated in a single ion trap.  Up to 14 ions in a string have now  successfully been put into an entangled state \cite{monz2011}; however, this number is approaching the limit of what is practically possible.  For larger numbers of ions it will be necessary to move to a system where ions are held in separate traps or in separate trapping regions within the same trap (see, for example, \cite{chiaverini2005}).  They would then be moved between different trapping regions in order to carry out gate operations between specific pairs of ions.  Transport of individual ions inside a trap structure without significant heating of the motional state has been   demonstrated in several experiments (e.g. \cite{blakestad2011}).

\subsection{Quantum simulation}  \label{sec:quantumsims}

\emph{Quantum simulation} refers to the use of one well-controlled quantum system to simulate the behaviour of another  quantum system that may be harder to control or measure.  Of course classical computers may be used for the purpose of simulating quantum mechanical systems (e.g. the behaviour of a set of spin--1/2 particles), but as the size of the system increases, simulation using a classical computer becomes computationally difficult and for large enough systems it is impossible to carry out meaningful simulations using current technology. The difficulty comes from having to keep track of all the coherences between individual elements of a quantum system.  It was Feynman who pointed out in 1982 that the best way to simulate such a system would be to use another quantum mechanical system \cite{feynman}.  It turns out that trapped ions are ideal for this type of application.  This is because the particles are very stable and well isolated from the environment, so the lifetimes of delicate quantum states such as entangled states and coherent states can be very long.  Furthermore, the particles have very well-controlled interactions with each other and with external fields such as magnetic fields, lasers and microwaves, and these interactions can be controlled to match the physics of different systems.  In addition, it is possible to measure the final internal (electronic) and external (motional) states of trapped ions to high accuracy using well-developed techniques, so detailed information on how the trapped-ion system is behaving can be obtained. Recent reviews of the field of quantum simulation with trapped ions  have been given by Schneider \emph{et al.} \cite{schneiderreview} and Johanning \emph{et al.} \cite{johanning2009}.  

Coulomb crystals are very important in this area because in general if trapped ions are to be used in some sort of quantum simulator, they need to be in a stable configuration with fixed distances between them, so an ICC is ideal for this.  

As an example, many experiments make use of a linear string of ions in a linear RF trap, where each ion qubit represents a spin--1/2 particle in a magnetic field.  In this case the ground state of the ion represents, for example, the spin down state $\vert\downarrow\rangle$ and the internal excited state of the ion represents the spin up state $|\uparrow\rangle$.  In order to simulate the well-known Ising model of interacting spins, we need to be able to simulate a magnetic field acting on the spins, and to engineer interactions between them.  

The effect of the magnetic field is represented by irradiating the ions with coherent radiation (a laser or radiofrequency field, depending on the type of qubits being employed).  This causes each qubit to oscillate continuously between the $\vert\downarrow\rangle$ and $\vert\uparrow\rangle$ states (an example of Rabi flopping), in the same way that a spin exposed to a perpendicular magnetic field would oscillate between its spin-down and spin-up states as it precessed around the magnetic field direction.  

The spin--spin interaction for two ions can be engineered using a standing wave light field which provides a force on each ion that depends on which state it is in.  In this way the force on one ion also leads to an effect on the other ion due to their mutual Coulomb interaction: if they are both in the same state, they move together but if they are in different states, they move in opposite directions and their Coulomb interaction energy changes.  This represents a spin-spin interaction because the energy of each ion now depends on the state of the other ion.  Alternatively, an effective spin--spin interaction can be engineered using a spatially varying magnetic field which causes ions to move slightly when their spin state changes, leading to an interaction between the ions mediated through the Coulomb interaction \cite{johanning2009}.

Experiments based on these ideas have been used to study the properties of a quantum magnet represented by two trapped ions \cite{friedenauer2008}.  By varying the parameters of the system, it was possible to observe both ferromagnetic and paramagnetic behaviour in this system.  By extending the system to three ions in a chain (the three-spin Ising model),  additional effects can be observed.  For this, the interactions are engineered such that adjacent spins have a lower energy when they point in opposite directions (i.e. anti-ferromagnet interactions).  Clearly not all three spins can arrange themselves in this way, so the system is said to exhibit \emph{spin frustration}.  This is closely linked to quantum mechanical entanglement in the system, and the well-controlled nature of ion trap techniques allows all the dynamics and properties of the system to be studied in detail \cite{kim2010}.

In a Penning trap it is possible to prepare a two-dimensional triangular array of ions, as discussed in Section \ref{sec:penningcrystals}.  This is ideal for simulation of a two-dimensional array of spins in a magnetic field \cite{bollingersims}.  In this experiment, performed on a planar crystal containing several hundred Be$^+$ ions (see Figure \ref{fig:hexagonal}), the ion qubit comprises the two Zeeman sub-levels of the ground state of the ion, separated by 124\,GHz in the magnetic field of the trap.  An optical dipole force that is dependent on the qubit state is generated using a pair of laser beams close in frequency to the laser cooling transition at 313\,nm, as described in Section \ref{sec:penningcrystals}.  This creates an effective interaction $J_{i,j}$ between any two qubits $i$ and $j$, which is mediated via the spatial properties of the transverse vibrational modes of the crystal \cite{bollingerdrumskin}.   By tuning the difference frequency of the laser beams, different vibrational modes can be selected, and this changes the way the effective spin--spin interaction depends on the separation of the qubits.  In general, it is found that $J_{i,j} \propto d_{i,j}^{-a}$ where $a$ is a constant and $d_{i,j}$ is the separation of qubits $i$ and $j$.  The spatial dependence of the interaction can be varied simply by changing the detuning of the laser beams, and this allows a range of interactions to be studied, from a constant force ($a=0$) through a Coulomb-like force ($a=1$) to a dipole-dipole force ($a=3$).

\subsection{Reaction cross-sections}  \label{sec:reactions}

Ions in Coulomb crystals can be used for investigation of cross-sections of various types of reaction \cite{drewsen2003}.   For instance, if a buffer gas is introduced into the ion trap chamber, any reactions between the buffer gas and the ions will result in a change in mass to charge ratio of the trapped ions; if they are in a Coulomb crystal, they will then separate from the original species, as described in Section \ref{sec:RFcrystals}.  This will result in a change in the shape of the crystal, which can be modelled through simulations.  Observation of the time-behaviour of the crystal shape, including the presence or absence of non-fluorescing species, will therefore give information on reaction cross sections.   The same principle was also used in the measurement of multi photon ionisation cross-sections of Mg and Mg$^+$.

A variation on this is to use a cloud of cold atoms rather than a buffer gas. In this case charge exchange cross sections or even the formation of cold molecular ions can be studied \cite{hall2013}.  For a review of this topic see \cite{CPcoldatoms}.

Experiments have shown that it is possible to create and trap simple molecular ions (e.g. H$_2^+$, HD$^+$) using electron bombardment of low pressure gas in an RF trap.  If the trap also contains laser-cooled Be$^+$ ions, the molecular ions are  sympathetically cooled to temperatures in the mK range \cite{blythe2005}.  One important potential application in this area is the  use of Coulomb crystals of sympathetically cooled molecular ions of hydrogen or deuterium for very precise measurement of ro-vibrational transition frequencies. The high precision and insensitivity to external perturbations could be exploited for frequency standards applications and also for the study of any possible variation of the fundamental constants with time \cite{karr2014, schiller2014}.

\subsection{Quantum Logic Spectroscopy} \label{sec:QLS}

It is perhaps questionable whether it is strictly correct to describe two trapped ions as a Coulomb crystal \cite{cricktwoions}, but we include it here in order to demonstrate a particularly important application for future frequency standards and ultra-high precision spectroscopy.  In order to create an optical frequency standard, ideally we require an ion that has an optical transition that is suitable for laser cooling, has a second transition (the \emph{clock transition}) that has very narrow linewidth, and whose electronic state can be detected efficiently by the presence or absence of fluorescence.  Some atomic ions that are good candidates for frequency standards applications (e.g. $^{27}$Al$^+$)  have suitable weak electronic transitions that are insensitive to external fields but do not have transitions that are suitable for laser cooling and for detection of the electronic state of the ion.  Quantum logic spectroscopy can couple such an ion to a second ion that can be laser cooled and whose electronic state can be detected with high efficiency (e.g. $^9$Be$^+$).  Introduced first by the Wineland group at NIST \cite{schmidt2005}, quantum logic spectroscopy uses the excitation of the vibrational mode of a two-ion Coulomb crystal to transfer information about the electronic state of one ion (the \emph{spectroscopy ion}) to the other ion (the \emph{logic ion}).  In this way, laser cooling and state detection are carried out on the logic ion, but the ultra-high resolution spectroscopy is carried out on the spectroscopy ion.  This allows a wider range of potential frequency standards candidates to be investigated.  

The technique has been  used by the Wineland group to demonstrate a frequency standard  with a fractional frequency inaccuracy below 10$^{-17}$ \cite{chou2010}.   It is also being exploited in other groups for the construction of ultra-high stability frequency standards and for the measurement of possible variations of the fundamental constants.  

\section{Conclusions}  \label{sec:conclusions}

In this review we have described the physics underlying the formation of ion Coulomb crystals, and we have discussed the methods used in various experiments to create, study and apply them in different areas of physics.  These novel and unique structures have properties that make them attractive for demonstration of a variety of physical phenomena, some of which are familiar, but less accessible, in other situations.  Their applications are growing  as scientists  further develop the techniques required to create them,  to  manipulate and measure their properties and to determine  the quantum state of the particles of which they are composed.  

Ion Coulomb crystals provide a very rich physical system for study, combining both classical physics in their formation and structure, and quantum physics in many of their applications.  The comparison with conventional crystals reveals both similarities and fundamental differences.  The range of applications, spanning across many areas of physics including thermodynamics, plasma physics, chemical reactions, fundamental physics, quantum optics and quantum information processing, demonstrates the power of ion traps and  laser cooling.  

Many of the studies that are performed with ion Coulomb crystals simply could not be carried out using alternative techniques, because the degree of precise control and delicate measurement which is possible with trapped ions cannot be achieved in other systems.  More applications are expected in the future for these uniquely beautiful and versatile objects.

\section*{Acknowledgements}
This work was supported in part by the European Commission STREP PICC (FP7 2007-2013
Grant number 249958). We also gratefully acknowledge financial support towards networking
activities from COST Action MP 1001 - Ion Traps for Tomorrows Applications. 

\section*{Biography}
Richard Thompson gained his DPhil at the University of Oxford in the area of atomic spectroscopy.  Following postdoctoral work in Germany and the UK, he  worked at the National Physical Laboratory on the first experiments with laser cooled trapped ions in the UK.  In 1986 he moved to Imperial College London where he is now a Professor of Experimental Physics.  His research has focussed on the dynamics of laser-cooled ions in Penning traps, including the creation and manipulation of ion Coulomb crystals in this system.

\bibliographystyle{tCPH}

\end{document}